\documentclass{article}
\usepackage{spconf,amsmath,graphicx}
\usepackage{booktabs}
\usepackage{mathrsfs}
\usepackage{caption}
\usepackage{mathtools}
\usepackage{amssymb}
\usepackage[utf8]{inputenc}
\usepackage{nccmath}
\usepackage[table]{xcolor}
\usepackage{multirow, multicol, makecell}
\usepackage{adjustbox}
\usepackage{mathrsfs}
\usepackage{enumerate}
\usepackage{enumitem}
\makeatletter
\def\blfootnote{\xdef\@thefnmark{}\@footnotetext}
\makeatother
\usepackage{scrextend}
\deffootnote[0em]{0em}{0em}
{\textsuperscript{\thefootnotemark}}
\usepackage{hyperref}
\usepackage{array}
\newcolumntype{C}[1]{>{\centering}p{#1}}
\newcolumntype{L}[1]{>{\PreserveBackslash\raggedright}p{#1}}
\usepackage{bm}
\DeclareMathOperator{\E}{\mathbb{E}}
\usepackage{setspace}

\title{Unsupervised representation learning for speaker recognition via Contrastive Equilibrium Learning}

\name{Sung Hwan Mun, Woo Hyun Kang, Min Hyun Han, Nam Soo Kim}
\address{Dept. of Electrical and Computer Engineering and INMC, Seoul National University, South Korea}

\begin{document}

%%%%%%%%%%%%%%%%%%%%%%%%%%%%%%%%%%%%%%%%%%%%%%%%%%%%%%%%%%%%%%%%%%%%%%%%%%%%%%%%%
\maketitle

%%%%%%%%%%%%%%%%%%%%%%%%%%%%%%%%%%%%%%%%%%%%%%%%%%%%%%%%%%%%%%%%%%%%%%%%%%%%%%%%%
\begin{abstract}
In this paper, we propose a simple but powerful unsupervised learning method for speaker recognition, namely Contrastive Equilibrium Learning (CEL), which increases the uncertainty on nuisance factors latent in the embeddings by employing the uniformity loss.
Also, to preserve speaker discriminability, a contrastive similarity loss function is used together. Experimental results showed that the proposed CEL significantly outperforms the state-of-the-art unsupervised speaker verification systems and the best performing model achieved 8.01\% and 4.01\% EER on VoxCeleb1 and VOiCES evaluation sets, respectively.
On top of that, the performance of the supervised speaker embedding networks trained with initial parameters pre-trained via CEL showed better performance than those trained with randomly initialized parameters.

\end{abstract}

\blfootnote{Our implementation is available here:}
\blfootnote{\texttt{github.com/msh9184/contrastive-equilibrium-learning}}

%%%%%%%%%%%%%%%%%%%%%%%%%%%%%%%%%%%%%%%%%%%%%%%%%%%%%%%%%%%%%%%%%%%%%%%%%%%%%%%%%

%%%%%%%%%%%%%%%%%%%%%%%%%%%%%%%%%%%%%%%%%%%%%%%%%%%%%%%%%%%%%%%%%%%%%%%%%%%%%%%%%
\section{Introduction}
\label{sec:intro}
Over the last decade, various deep learning-based speaker embedding techniques were developed for speaker recognition. Although these deep embedding techniques have shown outstanding performance on large scale datasets \cite{vox1, vox2, voices}, most of them are trained in a fully supervised manner \cite{xvector, gvlad, okabe, a-proto2, ge2e, margin}. However, since it is difficult to obtain a large amount of labeled data in real-life applications, learning speech representations without supervision is an important issue.

Recently, many efforts have been made to obtain good speech representations in an unsupervised learning manner.
In \cite{cpc, apc}, to learn the distribution of speech sequences, the speech representations were trained by using a probabilistic contrastive loss, which induces features to capture information useful to predict future samples.
In \cite{mock}, the multi-layer transformer encoders and the multi-head self-attention mechanism were employed to achieve bidirectional encoding for the speech representations.
\cite{lim, npc} proposed speech representations that capture the speaker identity maximizing the mutual information between the local embeddings extracted from the same utterance, which exploited the short-term active-speaker stationarity hypothesis to create contrastive samples from unlabeled data.

Maximizing the similarity between local features from the same utterance showed reasonable performance \cite{lim, npc}.
However, the segments cropped from the same sentence are likely to contain not only the same speaker identity but also the same nuisance factors (e.g., environment, noise, etc.).
This causes the embeddings to learn the shared nuisance attributes, which leads to low speaker identification or verification performance.
To alleviate this issue, authors in \cite{gcl, aat} proposed to augment the segments using different noises and Room Impulse Responses (RIR).
Furthermore, J. Huh, et al. \cite{aat} proposed an Augmentation Adversarial Training (AAT) strategy that penalizes the ability to predict the augmentation types so that the embeddings can be optimized to be channel-invariant.
Although the AAT shows meaningful improvement over previous works on unsupervised representation learning in terms of speaker verification, optimizing the adversarial loss using the gradient reversal layer is known to be very unstable and sensitive to hyper-parameter setting \cite{unstable2, kang}.
%%%%%%%%%%%%%%%%%%%%%%%%%%%%%%%%%%%%%%%%%%%%%%%%%%%%%%%%%%

In this paper, we propose a simple but powerful training strategy, Contrastive Equilibrium Learning (CEL).
Unlike the conventional techniques, the proposed CEL increases the uncertainty on nuisance factors latent in the embeddings by employing the uniformity loss.
Minimizing the uniformity loss forces the embeddings to be in the equilibrium state (i.e., uniformly distributed over the unit hyper-sphere), which leads to having the highest entropy \cite{uni-ent}. In order to formulate the uniformity loss, we exploit the total pairwise potential calculated with a Gaussian kernel function similarly to \cite{kernel1, kernel2, ali-uni}.

However, optimizing only the uniformity loss may also increase the uncertainty on the speaker's information inherent in the embeddings. In order to preserve speaker discriminability, we use the similarity loss function together. The similarity loss functions are formulated using the end-to-end contrastive learning-based objectives.

Experimental results show that the proposed method CEL significantly outperforms the state-of-the-art unsupervised speaker verification systems on VoxCeleb1 test and VOiCES evaluation set.
On top of that, the performance of the supervised speaker embedding networks trained with initial parameters pre-trained via CEL showed better performance than those trained with randomly initialized parameters. Through this experimental result, we demonstrated that the proposed CEL can be used to find good initial parameters for the conventional deep speaker embedding systems.

%%%%%%%%%%%%%%%%%%%%%%%%%%%%%%%%%%%%%%%%%%%%%%%%%%%%%%%%%%%%%%%%%%%%%%%%%%%%%%%%%
\section{Proposed Method}
The overall process of the proposed method CEL is shown in Fig. 1. To train the speaker embedding network in an unsupervised learning fashion, we first randomly crop two segments from a single utterance and individually apply augmentations with different additive noises and reverberations. The speaker embeddings are extracted from the front-end encoder and normalized to be on the unit hyper-sphere.

Assuming that there are no overlapping speakers within a batch, we train the front-end encoder via the proposed CEL strategy, which consists of uniformity loss and similarity loss.
 
%%%%%%%%%%%%%%%%%%%%%%%%%%%%%%%%%%%%%%%%%%%%%%%%%%%%%%%%%%%%%%%%%%%%%%%%%%%%%%%%%
\subsection{Loss functions}
In the proposed CEL framework, a batch is comprised of total \textit{$2K$}-segments $\{(\textbf{x}_{i,1},\textbf{x}_{i,2})\}_{i=1}^K$ cropped from \textit{$K$}-utterances $\{\textbf{x}_{i}\}_{i=1}^K$. Given $f:\mathbb{R}^{n\times T} \rightarrow \mathbb{R}^{m}$ which is the front-end encoder network that maps the speech segments of dimension $n$ with the frame length $T$ to $\ell_2$-normalized embeddings of dimension $m$, $\{(f(\tilde{\textbf{x}}_{i,1}),f(\hat{\textbf{x}}_{i,2}))\}_{i=1}^K$ denotes positive pairs of two speaker embeddings with different augmentations.

%%%%%%%%%%%%%%%%%%%%%%%%%%%%%%%%%%%%%%%%%%%%%%%%%%%%%%%%%%%%%%%%%%%%%%%%%%%%%%%%%
\vfill
\noindent\textbf{Uniformity loss.}
In order to force the embeddings to reach an equilibrium state, namely a state of minimal energy (i.e., the distribution optimizing this metric should converge to uniform distribution on the hyper-sphere), we leverage the pairwise Gaussian potential kernel known as the Radial Basis Function (RBF) kernel, $\textbf{G}_{t}: \mathbb{R}^{m} \times \mathbb{R}^{m}\rightarrow \mathbb{R_+}$

%%%%%%%%%%%%%%%%%%%%%%%%%%%%%%%%%%%%%%%%%%%%%%%%%%%%%%%%%%%%%%%%%%%%%%%%%%%%%%%%%
\begin{equation}
\label{eq4}
\begin{split}
  \textbf{G}_t(f(\textbf{x}_i), f(\textbf{x}_j)) \triangleq 
  \exp\left(-t {\left\lVert f(\textbf{x}_i)- f(\textbf{x}_j) \right\rVert}^2_2\right)
\end{split}
\end{equation}

%%%%%%%%%%%%%%%%%%%%%%%%%%%%%%%%%%%%%%%%%%%%%%%%%%%%%%%%%%%%%%%%%%%%%%%%%%%%%%%%%
\noindent where $t > 0$ is a fixed parameter. Similarly to \cite{kernel1, kernel2, ali-uni}, the uniformity loss is defined as the logarithm of the average pairwise Gaussian potential as follows:

%%%%%%%%%%%%%%%%%%%%%%%%%%%%%%%%%%%%%%%%%%%%%%%%%%%%%%%%%%%%%%%%%%%%%%%%%%%%%%%%%
\begin{equation}
\label{eq5}
\begin{split}
  \mathcal{L}_{\text u} & \triangleq \log \displaystyle \E_{\textbf{x}_i, \textbf{x}_j \overset{\text{\tiny{i.i.d.}}}{\sim}  p_{\texttt{\tiny data}}} \left[\textbf{G}_t(f(\textbf{x}_i), f(\textbf{x}_j))\right] \\
  & = \log \displaystyle \E_{\textbf{x}_i, \textbf{x}_j \overset{\text{\tiny{i.i.d.}}}{\sim}  p_{\texttt{\tiny data}}} \left[\exp \left(-t {\left\lVert f(\textbf{x}_i)- f(\textbf{x}_j) \right\rVert}^2_2\right)\right]
\end{split}
\end{equation}

%%%%%%%%%%%%%%%%%%%%%%%%%%%%%%%%%%%%%%%%%%%%%%%%%%%%%%%%%%%%%%%%%%%%%%%%%%%%%%%%%
\noindent where minimizing equation (2) leads the embedding vectors to have on uniform distribution \cite{kernel2}.

Analogous to \cite{ali-uni}, the uniformity loss within a batch can be calculated by considering each pair of $\{\tilde{\textbf{x}}_{i,1}\}_{i}$ and $\{\tilde{\textbf{x}}_{i,2}\}_{i}$:

%%%%%%%%%%%%%%%%%%%%%%%%%%%%%%%%%%%%%%%%%%%%%%%%%%%%%%%%%%%%%%%%%%%%%%%%%%%%%%%%%
\begin{equation}
\label{eq6}
\begin{split}
  \mathcal{L}_{\text u} = {1 \over 2} \log {\left( {1 \over \binom{K}{2}} \sum_{i \neq j}{ \exp (-t {\left\lVert f(\tilde{\textbf x}_{i,1})-f(\hat{\textbf x}_{j,1}) \right\rVert}^2_2}) \right)}
  \\ + {1 \over 2} \log {\left( {1 \over \binom{K}{2}} \sum_{i \neq j}{ \exp (-t {\left\lVert f(\tilde{\textbf x}_{i,2})-f(\hat{\textbf x}_{j,2}) \right\rVert}^2_2)} \right)}.
\end{split}
\end{equation}

%%%%%%%%%%%%%%%%%%%%%%%%%%%%%%%%%%%%%%%%%%%%%%%%%%%%%%%%%%%%%%%%%%%%%%%%%%%%%%%%%
\noindent\textbf{Similarity loss}.
In order to ensure that the embeddings of the positive pairs to be similar, while pushing the embeddings extracted from the negative pairs apart, we exploit the angular prototypical loss proposed in \cite{a-proto2}. For our unsupervised learning settings, the angular prototypical similarity loss is formulated as follows:

%%%%%%%%%%%%%%%%%%%%%%%%%%%%%%%%%%%%%%%%%%%%%%%%%%%%%%%%%%%%%%%%%%%%%%%%%%%%%%%%%
\begin{equation}
  \label{eq1}
  \mathcal{L}_{\text {s-ap}} = - {1 \over K} \sum_{i=1}^{K} \log { \exp{(\textbf{S}_{\theta}(f(\tilde{\textbf x}_{i,1}),f(\hat{\textbf x}_{i,2})))}  \over {\sum_{j=1}^{K} \exp {(\textbf{S}_{\theta}(f(\tilde{\textbf x}_{i,1}),f(\hat{\textbf x}_{j,2})))} } },
\end{equation}

%%%%%%%%%%%%%%%%%%%%%%%%%%%%%%%%%%%%%%%%%%%%%%%%%%%%%%%%%%%%%%%%%%%%%%%%%%%%%%%%%
\begin{equation}
  \label{eq2}
  \textbf{S}_{\theta}(f(\tilde{\textbf x}_{i,1}), f(\hat{\textbf x}_{j,2})) = 
  w {{f(\tilde{\textbf x}_{i,1})^T f(\hat{\textbf x}_{j,2})} \over 
  {\left\lVert{f(\tilde{\textbf x}_{i,1})}\right\rVert \left\lVert{f(\hat{\textbf x}_{j,2})}\right\rVert}} + b
\end{equation}

%%%%%%%%%%%%%%%%%%%%%%%%%%%%%%%%%%%%%%%%%%%%%%%%%%%%%%%%%%%%%%%%%%%%%%%%%%%%%%%%%
\noindent where $\textbf{S}_{\theta}: \mathbb{R}^{m} \times \mathbb{R}^{m}\rightarrow \mathbb{R}$ is the affine transformation of the cosine similarity between two speaker embeddings of dimension $m$, $w$ and $b$ $\in \theta$ are trainable parameters for scale and bias respectively.
Additionally, we take the following angular contrastive similarity loss into account:
%%%%%%%%%%%%%%%%%%%%%%%%%%%%%%%%%%%%%%%%%%%%%%%%%%%%%%%%%%%%%%%%%%%%%%%%%%%%%%%%%
\begin{equation}
\label{eq3}
\begin{split}
  \mathcal{L}_{\text{s-ac}} = -{1 \over 2K} \sum_{i=1}^{K} \log { \exp{(\textbf{S}_{\theta}(f(\tilde{\textbf x}_{i,1}),f(\hat{\textbf x}_{i,2})))}  \over 
  {\sum_{j=1}^{K} \exp{(\textbf{S}_{\theta}(f(\tilde{\textbf x}_{i,1}),f(\hat{\textbf x}_{j,2})))} } } 
 \\ - {1 \over 2K} \sum_{i=1}^{K} \log { \exp{(\textbf{S}_{\theta}(f(\tilde{\textbf x}_{i,1}),f(\hat{\textbf x}_{i,2})))}  \over 
 {\sum_{j=1}^{K} \exp{(\textbf{S}_{\theta}(f(\tilde{\textbf x}_{j,1}),f(\hat{\textbf x}_{i,2})))} } }.
\end{split}
\end{equation}

%%%%%%%%%%%%%%%%%%%%%%%%%%%%%%%%%%%%%%%%%%%%%%%%%%%%%%%%%%%%%%%%%%%%%%%%%%%%%%%%%
 This loss function has been shown effective in many recent representation learning methods \cite{cpc, moco}.
 
%%%%%%%%%%%%%%%%%%%%%%%%%%%%%%%%%%%%%%%%%%%%%%%%%%%%%%%%%%%%%%%%%%%%%%%%%%%%%%%%%
\begin{figure}[t]
\begin{minipage}[b]{1.0\linewidth}
  \centering
  % candidates {0927_2, 0927_11, 0927_9, 0927_10}
  \centerline{\includegraphics[width=8.8cm]{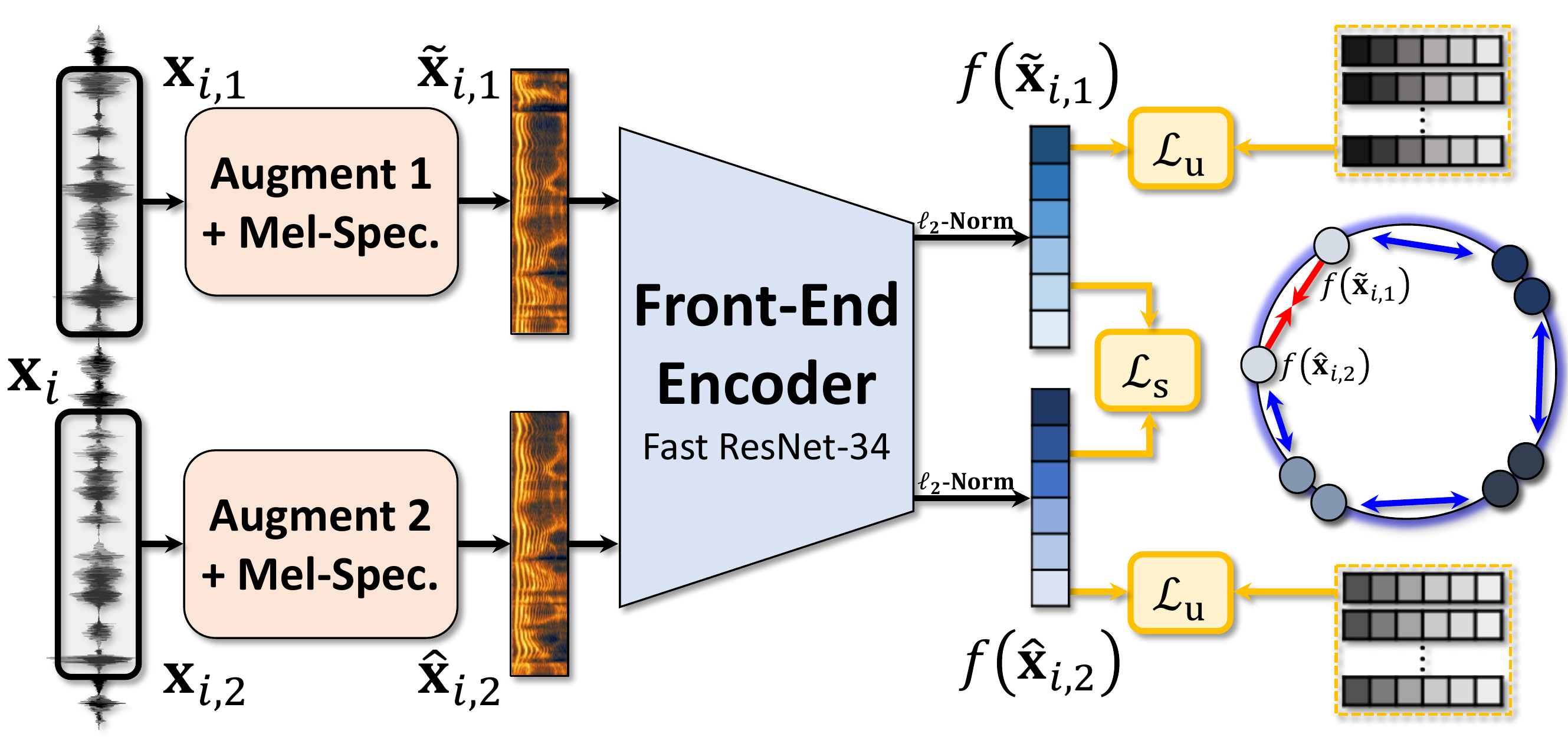}}
\vspace{0cm}
\end{minipage}
\caption{Process of Contrastive Equilibrium Learning (CEL)}
\label{fig:res}
\end{figure}

%%%%%%%%%%%%%%%%%%%%%%%%%%%%%%%%%%%%%%%%%%%%%%%%%%%%%%%%%%%%%%%%%%%%%%%%%%%%%%%%%

The front-end encoder network is trained using the two objective functions, i.e., the combination of uniformity loss with a weighting factor $\lambda$ and similarity loss as follows:

%%%%%%%%%%%%%%%%%%%%%%%%%%%%%%%%%%%%%%%%%%%%%%%%%%%%%%%%%%%%%%%%%%%%%%%%%%%%%%%%%
\begin{equation}
\label{eq7}
\begin{split}
  \mathcal{L}_{\text{total}} = \lambda \mathcal{L}_{\text u} + \mathcal{L}_{\text s}.
\end{split}
\end{equation}

%%%%%%%%%%%%%%%%%%%%%%%%%%%%%%%%%%%%%%%%%%%%%%%%%%%%%%%%%%%%%%%%%%%%%%%%%%%%%%%%%
\section{Experiments}
\label{sec:pagestyle}

\subsection{Unsupervised learning}

\noindent\textbf{Experimental setups.}
During training, we randomly cropped the input utterance to two 180-frames segments with 10-ms window hop-size, and two crops were differently added MUSAN noises \cite{musan} and convoluted with finite RIR filters as in \cite{aat}.
The Fast ResNet-34 proposed in \cite{a-proto2} was used as the front-end encoder, and log Mel-spectrograms of dimension 40 were extracted as acoustic features.
In all experiments, the constant factor for pairwise Gaussian kernel $\textbf{G}_t$ was fixed to $t$$=$2.
The networks were trained in an unsupervised manner on the development set of VoxCeleb2 \cite{vox2} with 1,092,009 utterances and evaluated on the original test set of VoxCeleb1 \cite{vox1}. We also use VOiCES 2019 Challenge development set according to the trial list provided by \cite{voices} to verify the generalization capacity of the experimented systems on the out-of-domain data.

The benchmarks contain Disent. \cite{disent}, CDDL \cite{cddl}, GCL \cite{gcl}, I-vector, and AAT with angular prototypical loss \cite{aat}. Disent. and CDDL employed the cross-modal synchrony between faces and audio in a video for learning the embeddings in a self-supervised fashion. 
GCL and AAT exploited data augmentation using different noises and RIRs in an unsupervised manner.
Furthermore, AAT used adversarial training so that the embeddings can be learned to be channel-invariant.

The networks were trained using Adam optimizer with an initial learning rate of 0.001, decreased by 5\% per 10-epochs.
The models were learned using an NVIDIA Tesla M40 single GPU with 24GB memory for 500-epochs.
For the back-end scoring, the cosine similarity-based technique used in \cite{vox2, a-proto2, aat} was taken, and Equal Error Rate (EER) was evaluated as the performance measure.

%%%%%%%%%%%%%%%%%%%%%%%%%%%%%%%%%%%%%%%%%%%%%%%%%%%%%%%%%%%%%%%%%%%%%%%%%%%%%%%%%
\smallskip \noindent\textbf{Results.}
We first compare the models trained via the proposed CEL with the previous benchmarks on unsupervised representation learning shown in Table 1, where the parameters were fixed to $\lambda$$=$1, $K$$=$200.
We experimented with two proposed models using either angular prototypical (A-Prot) or angular contrastive (A-Cont) loss as similarity loss together with uniformity loss (Unif).
The results on both VoxCeleb1 and VOiCES evaluation sets showed that our models outperformed the conventional works.
The best performing model, which was trained with Unif and A-Prot loss, achieved 8.01\% and 4.01\% in terms of EER on VoxCeleb1 and VOiCES evaluation sets, respectively.

%%%%%%%%%%%%%%%%%%%%%%%%%%%%%%%%%%%%%%%%%%%%%%%%%%%%%%%%%%%%%%%%%%%%%%%%%%%%%%%%%
We further conduct the ablation study on the VoxCeleb1 original test set to demonstrate the the effectiveness of each hyper-parameter on the speaker verification performance shown in Table 2. The two components were analyzed for the ablation study: the batch size $K$ and the uniformity weight $\lambda$.
In these experiments, A-Prot loss was used as a similarity loss in common.
Usually, the contrastive objectives are known to benefit from larger batch size \cite{a-proto2}. However, since we assume that every utterance contains only one person’s speech, increasing the batch size may have a negative impact \cite{aat}. This is shown in the results of Table 2 where the best performance was observed in $K$$=$200, which outperformed larger batch sizes.
Also, as shown in Table 2, we observed that the equal-weighted summation (i.e., $\lambda$=1) between uniformity and similarity loss is the best performing point in our settings.

%%%%%%%%%%%%%%%%%%%%%%%%%%%%%%%%%%% Table 1 %%%%%%%%%%%%%%%%%%%%%%%%%%%%%%%%%%%
\begin{table}[t]
\centering
\caption{Speaker verification results of unsupervised learning}
\label{tab1:table}
%\renewcommand{\arraystretch}{0.9}
%\renewcommand{\tabcolsep}{1.55mm}
%\resizebox{\columnwidth}{!}{
%\normalsize{
\small{
%\footnotesize{
\begin{tabular}{lcrr} %L{2.85cm} C{0.5cm}
\toprule[.1em]
      \multirow{2}{*}{\textbf{Model}}
    & \multirow{2}{*}{\textbf{Aug.}}
    & \multicolumn{1}{r}{\textbf{VoxCeleb1}}
    & \multicolumn{1}{r}{\textbf{VOiCES}} \\
        %\cmidrule(lr){3-4}  %\cmidrule(lr){5-6}
        &
        & \textbf{EER[\%]}
        & \textbf{EER[\%]} \\
\midrule[.05em]
      Disent. \cite{disent}
    & --
    & 22.09
    & -- \\
      CDDL \cite{cddl}
    & --
    & 17.52
    & -- \\
      GCL \cite{gcl}
    & \checkmark 
    & 15.26
    & -- \\ 
      I-vector \cite{aat}
    & --
    & 15.28
    & 17.49 \\ 
      %\multirow{2}{*}{Proto. \cite{aat}} 
      Prot \cite{aat}
    & -- 
    & 27.30 
    & 29.69 \\ 
      Prot \cite{aat}
    & \checkmark
    & 10.16 
    & 5.82 \\ 
      %\multirow{2}{*}{A-Proto. \cite{aat}} 
      A-Prot \cite{aat}
    & -- 
    & 25.37 
    & 32.21 \\ 
      A-Prot \cite{aat}
    & \checkmark 
    & 9.56  
    & 5.65 \\ 
      AAT + Prot \cite{aat} 
    & \checkmark 
    & 9.36
    & 5.26 \\ 
      AAT + A-Prot \cite{aat} 
    & \checkmark 
    & 8.65 
    & 4.96 \\ 
    \arrayrulecolor{black!40}\midrule[.04em]
      \textbf{Unif + A-Prot (CEL)}
    & \checkmark
    & \textbf{8.01}
    & \textbf{4.01} \\ 
      \textbf{Unif + A-Cont (CEL)} 
    & \checkmark
    & \textbf{8.05} 
    & \textbf{4.69} \\ 
    \addlinespace[.0em]
\arrayrulecolor{black}\bottomrule[.1em]
\end{tabular}
%}
}
\end{table}

%%%%%%%%%%%%%%%%%%%%%%%%%%%%%%%%%% Table 2 %%%%%%%%%%%%%%%%%%%%%%%%%%%%%%%%%%%
\begin{table}[t]
\centering
\caption{Ablation study on VoxCeleb1 original test set}
\label{tab2:table}

\renewcommand{\tabcolsep}{2.8mm} % 3.8mm
%\normalsize{
\small{
\begin{tabular}{cccc}
\toprule[.1em]
    \textbf{Ablation}
    & \textbf{Batch size}
    & \textbf{Unif weight}
    & \textbf{EER[\%]} \\
\midrule[.05em]
    % batch size
      \multirow{4}{*}{Batch size} 
    & $K$ $=$ \text{200}
    & $\lambda$ $=$ \text{1.0}
    & \textbf{8.01} \\
    & $K$ $=$ \text{300}
    & $\lambda$ $=$ \text{1.0}
    & 8.25 \\
    & $K$ $=$ \text{500}
    & $\lambda$ $=$ \text{1.0}
    & 8.22 \\
    & $K$ $=$ \text{800}
    & $\lambda$ $=$ \text{1.0}
    & 8.12 \\
\arrayrulecolor{black!}\midrule[.05em]
    % loss weight
    \multirow{3}{*}{\makecell[c]{Unif weight}}
        & $K$ $=$ \text{200}
        & $\lambda$ $=$ \text{0.5}
        & 8.72 \\
        & $K$ $=$ \text{200}
        & $\lambda$ $=$ \text{1.0}
        & \textbf{8.01} \\
        & $K$ $=$ \text{200}
        & $\lambda$ $=$ \text{2.0}
        & 8.20 \\
\arrayrulecolor{black}\bottomrule[.1em]
\end{tabular}
%}
}
\end{table}

%%%%%%%%%%%%%%%%%%%%%%%%%%%%%%%%%% Table 3 %%%%%%%%%%%%%%%%%%%%%%%%%%%%%%%%%%%
\begin{table*}[t]
\centering
\caption{The speaker verification results using the parameters either randomly initialized or pre-trained via CEL}
\label{tab3:table}
%\renewcommand{\arraystretch}{1.05}
%\renewcommand{\tabcolsep}{2.3mm}
%\renewcommand{\tabcolsep}{2.3mm} % 4.55mm
%\normalsize{
\small{
\begin{tabular}{cccccccc} %L{2.85cm} C{0.5cm}
\toprule[.1em]
    \multicolumn{2}{c}{\multirow{1}{*}{\makecell[c]{\textbf{Unsupervised} \textbf{pre-training}}}}
    & \multicolumn{2}{c}{\multirow{1}{*}{\makecell[c]{\textbf{Supervised} \textbf{fine-tunning}}}} 
    & \multicolumn{2}{c}{\makecell[c]{\textbf{VoxCeleb1}}}
    & \multicolumn{2}{c}{\makecell[c]{\textbf{VOiCES}}} \\
        \cmidrule(lr){1-2} \cmidrule(lr){3-4} \cmidrule(lr){5-6} \cmidrule(lr){7-8}
        \textbf{Dataset}
        & \textbf{Objective}
        & \textbf{Dataset}
        & \textbf{Objective}
        & \textbf{EER[\%]}
        & \textbf{MinDCF}
        & \textbf{EER[\%]}
        & \textbf{MinDCF} \\
\midrule[.05em]
    \multicolumn{2}{c}{\multirow{6}{*}{\makecell[c]{Random \\ Initialization}}}
    & \multirow{6}{*}{\makecell[c]{VoxCeleb1 \\ w/ labels}}
    & A-Prot
    & 5.32
    & 0.3792
    & 6.58
    & 0.4852 \\
    &
    &
    & A-Cont
    & 5.21
    & 0.3750
    & 6.97
    & 0.4909 \\
    & 
    & 
    & GE2E
    & 5.98 
    & 0.4258
    & 7.75
    & 0.6182 \\
    & 
    & 
    & CosFace
    & 5.51
    & 0.3402 
    & 7.45
    & 0.4494 \\
    & 
    & 
    & ArcFace
    & 5.53
    & 0.3516
    & 6.68
    & 0.4944 \\
    & 
    & 
    & AdaCos
    & 5.89
    & 0.4047
    & 7.26
    & 0.5820 \\
\arrayrulecolor{black!40}\midrule[.04em]
    \multirow{6}{*}{\makecell[c]{\textbf{VoxCeleb2} \\ \textbf{w/o labels}}}
    & \multirow{6}{*}{\makecell[c]{Unif+A-Prot}} %$\mathcal{L}_{\text{s-ap}}+\mathcal{L}_{\text u}$
    & \multirow{6}{*}{\makecell[c]{\textbf{VoxCeleb1} \\ \textbf{w/ labels}}}
    & \textbf{A-Prot}
    & \textbf{2.33}
    & \textbf{0.1741}
    & 2.78
    & 0.2110 \\
    & 
    & 
    & A-Cont
    & 2.35
    & 0.1804
    & 2.69
    & 0.2138 \\    
    & 
    & 
    & \textbf{GE2E}
    & 2.52
    & 0.1876
    & \textbf{2.59}
    & \textbf{0.1906} \\
    & 
    & 
    & CosFace
    & 2.83
    & 0.1947
    & 2.74
    & 0.2081 \\
    & 
    & 
    & ArcFace
    & 2.84
    & 0.1797
    & 2.89
    & 0.2148 \\ 
    & 
    & 
    & AdaCos
    & 2.76
    & 0.2061
    & 2.82
    & 0.2438 \\
\arrayrulecolor{black!40}\midrule[.04em]
    \multirow{6}{*}{\makecell[c]{\textbf{VoxCeleb2} \\ \textbf{w/o labels}}}
    & \multirow{6}{*}{\makecell[c]{Unif+A-Cont}} %$\mathcal{L}_{\text{s-ac}}+\mathcal{L}_{\text u}$
    & \multirow{6}{*}{\makecell[c]{\textbf{VoxCeleb1} \\ \textbf{w/ labels}}}
    & A-Prot
    & 2.42
    & 0.1774
    & 2.65
    & 0.2107 \\
    & 
    & 
    & \textbf{A-Cont}
    & \textbf{2.36}
    & \textbf{0.1766}
    & 2.63
    & 0.2048 \\    
    & 
    & 
    & \textbf{GE2E}
    & 2.63
    & 0.1917
    & \textbf{2.39}
    & \textbf{0.1806} \\
    & 
    & 
    & CosFace
    & 2.84
    & 0.1871
    & 3.03
    & 0.2323 \\
    & 
    & 
    & ArcFace
    & 2.84
    & 0.1979
    & 2.64
    & 0.2171 \\ 
    & 
    & 
    & AdaCos
    & 2.79
    & 0.2225
    & 2.96
    & 0.2553 \\    
\arrayrulecolor{black}\bottomrule[.1em]
\end{tabular}
%}
}
\end{table*}

%%%%%%%%%%%%%%%%%%%%%%%%%%%%%%%%%%%%%%%%%%%%%%%%%%%%%%%%%%%%%%%%%%%%%%%%%%%%%%%%%
\subsection{Fine-tuning using the models pre-trained via CEL}

\noindent\textbf{Experimental setups.}
The front-end encoders were pre-trained as in Section 3.1 using the proposed CEL technique. We leveraged two pre-trained models using either A-Prot or A-Cont loss together with Unif loss where $\lambda$=1, $K$=200. 
To fine-tune the speaker embedding networks in a supervised manner, the development set of VoxCeleb1 with 148,642 utterances from 1,211 speakers was used.
We used a 300-frames segment from randomly sampled utterances without data augmentation.
Following objective functions were employed for fine-tuning the networks:
\begin{itemize}[noitemsep, leftmargin=*, label=\scalebox{.7}{\textbullet}, topsep=4.0pt] %, itemsep=1.0pt, partopsep=1.0pt, parsep=1.0pt]
\item A-Prot \cite{a-proto2}: Angle prototypical loss given in equation (1),
\item A-Cont: Angle contrastive loss given in equation (3),
\item GE2E \cite{ge2e}: Generalized end-to-end loss based on the cosine similarity with the learnable parameters,
\item CosFace \cite{cosface, cosface2}: Additive margin softmax loss also called \textit{AM-softmax loss}. In this experiments, we set the cosine margin $m$$=$0.2 and scale factor $s$$=$30,
\item ArcFace \cite{arcface}: Additive angular margin softmax loss also called \textit{AAM-softmax loss}. In this experiments, we set additive angular margin $m$$=$0.2 and scale factor $s$$=$30,
\item AdaCos \cite{adacos}: Adaptive scaling cosine loss proposed in \cite{adacos}. We used the loss with a dynamic scale parameter.
\end{itemize}

The Adam optimizer with an initial learning rate of 0.001, reduced by 10\% per 10-epochs, was used for 250-epochs, and we trained the models using the same GPU and back-end scoring method as those in section 3-1.
Two measures were analyzed, which are EER and a Minimum Detection Cost Function (MinDCF).
The parameters of MinDCF were set as $\mathcal{C}_{miss}$$=$1, $\mathcal{C}_{fa}$$=$1 and $\mathcal{P}_{target}$$=$0.05.

%%%%%%%%%%%%%%%%%%%%%%%%%%%%%%%%%%%%%%%%%%%%%%%%%%%%%%%%%%%%%%%%%%%%%%%%%%%

\smallskip\noindent\textbf{Results.}
The speaker verification results using the parameters either randomly initialized or pre-trained via CEL are given in Table 3.
All systems fine-tuned with the initial parameters pre-trained via CEL showed much better performance than those trained with randomly initialized weights.
These results demonstrated that the CEL can be used to find good initial parameters for the conventional deep speaker embedding systems.
Moreover, the contrastive objectives-based models showed better performances compared to the softmax-based models.
The best performances were 2.33\% and 2.39\% in terms of EER on VoxCeleb1 and VOiCES sets, respectively.

The comparison of the previous works and our models trained with the initial parameters pre-trained via CEL is shown in Table 4. Experimental results showed remarkable improvement over the conventional benchmarks. When using the VoxCeleb1 development set for fine-tuning, the proposed model showed the EER of 2.33\%. By employing the larger datasets VoxCeleb2 and VoxCeleb1\&2, we obtained 2.05\% and 1.81\% in terms of EER, respectively. These results outperform the performaces of state-of-the-art methods.

%%%%%%%%%%%%%%%%%%%%% Table 4 %%%%%%%%%%%%%%%%%%%%%%%%%%%
\begin{table}[t]
\centering
\caption{Comparison of previous works and proposed models}
\label{tab4:table}
\renewcommand{\tabcolsep}{1.2mm}
%\normalsize{
\small{
\begin{tabular}{lcccr}
\toprule[.1em]
    \textbf{Model}
    & \textbf{\makecell{Training set}}
    & \textbf{Back-end}
    & \textbf{EER} \\
\midrule[.05em]
      Nagrani et al. \cite{vox1}
    %& \multirow{6}{*}{\makecell[c]{VoxCeleb1}}
    & VoxCeleb1
    & PLDA
    & 8.80 \\
      Ravanelli \& Bengio \cite{lim}
    & VoxCeleb1
    & Cosine
    & 5.80 \\
      Han et al. \cite{han}
    & VoxCeleb1
    & PLDA
    & 5.11 \\
      Kang et al. \cite{kang}
    & VoxCeleb1
    & PLDA
    & 4.40 \\
      Okabe et al. \cite{okabe}
    & VoxCeleb1
    & Cosine
    & 3.85 \\
\arrayrulecolor{black!40}\midrule[.04em]
    Xie et al. \cite{gvlad}
    & VoxCeleb2
    & Cosine
    & 3.22 \\
    Xiang et al. \cite{margin}
    & VoxCeleb2
    & Cosine
    & 2.69 \\    
      Kaldi recipe \cite{monteiro}
    & VoxCeleb2
    & PLDA
    & 2.51 \\
      Monteiro et al. \cite{monteiro}
    & VoxCeleb2
    & LRD-E2E
    & 2.51 \\
      Chung et al. \cite{a-proto2}
    & VoxCeleb2
    & Cosine
    & 2.21 \\
\arrayrulecolor{black!40}\midrule[.04em]
      \textbf{Ours \scriptsize{(Unif + A-Prot}}$\rightarrow$\textbf{\scriptsize{A-Prot})}
    & \textbf{VoxCeleb1}
    & \textbf{Cosine} %$\ell_2$ \textbf{Distance} 
    & \textbf{2.33} \\
      \textbf{Ours \scriptsize{(Uni + A-Cont}}$\rightarrow$ \textbf{\scriptsize{ArcFace)}}
    & \textbf{VoxCeleb2}
    & \textbf{Cosine} %$\ell_2$ \textbf{Distance} 
    & \textbf{2.05} \\
      \textbf{Ours \scriptsize{(Unif + A-Cont}}$\rightarrow$ \textbf{\scriptsize{GE2E)}}
    & \textbf{VoxCeleb1\&2}
    & \textbf{Cosine} %$\ell_2$ \textbf{Distance} 
    & \textbf{1.81} \\
\arrayrulecolor{black}\bottomrule[.1em]
\end{tabular}
%}
}
\end{table}

%%%%%%%%%%%%%%%%%%%%%%%%%%%%%%%%%%%%%%%%%%%%%%%%%%%%%%%%%%%%
\section{Conclusion}
\label{sec:typestyle}
In this work, we introduced a simple but powerful training strategy, namely Contrastive Equilibrium Learning.
The proposed CEL increases the uncertainty on nuisance factors latent in the embeddings by employing the uniformity loss. Additionally, to preserve speaker discriminability, the similarity losses is used together.
Experimental results showed that the proposed CEL significantly outperforms the state-of-the-art unsupervised speaker verification systems on VoxCeleb1 and VOiCES sets.
On top of that, the performance of the supervised speaker embedding networks trained with initial parameters pre-trained via CEL showed better performance than those trained with randomly initialized parameters.

\begin{center}
    \vspace{0.1cm}
    \color{black!40}
    \hrule height 0.015cm
    \vspace{-0.35cm}
\end{center}
\noindent \footnotesize{\textbf{Acknowledgements.} This research was supported and funded by the Korean National Police Agency. [Project Name: Real-time speaker recognition via voiceprint analysis / Project Number: PA-J000001-2017-101]}

\begingroup
\setstretch{0.8}
\normalsize
\bibliographystyle{IEEEbib}
\bibliography{strings,refs}
\endgroup

\end{document}